# On the performance of large monolithic LaCl$_3$(Ce) crystals coupled to pixelated silicon photosensors

P. Olleros[a,*], L. Caballero[a,†], C. Domingo-Pardo[a], V. Babiano[a], I. Ladarescu[a], D. Calvo[a], P. Gramage[a], E. Nacher[b], J.L. Tain[a], A. Tolosa[a]

[a] *Instituto de Física Corpuscular (CSIC-University of Valencia), C/Catedrático José Beltrán, 2 46980 Paterna, Spain*

[b] *Instituto de Estructura de la Materia, CSIC, Madrid, Spain*
   E-mail: `Luis.Caballero@ific.uv.es`

ABSTRACT: We investigate the performance of large area radiation detectors, with high energy- and spatial-resolution, intended for the development of a Total Energy Detector with gamma-ray imaging capability, so-called i-TED. This new development aims for an enhancement in detection sensitivity in time-of-flight neutron capture measurements, versus the commonly used C$_6$D$_6$ liquid scintillation total-energy detectors. In this work, we study in detail the impact of the readout photosensor on the energy response of large area (50×50 mm$^2$) monolithic LaCl$_3$(Ce) crystals, in particular when replacing a conventional mono-cathode photomultiplier tube by an 8×8 pixelated silicon photomultiplier. Using the largest commercially available monolithic SiPM array (25 cm2), with a pixel size of 6×6 mm$^2$, we have measured an average energy resolution of 3.92% FWHM at 662 keV for crystal thicknesses of 10, 20 and 30 mm. The results are confronted with detailed Monte Carlo (MC) calculations, where both optical processes and properties have been included for the reliable tracking of the scintillation photons. After the experimental validation of the MC model, se use our MC code to explore the impact of different a photosensor segmentation (pixel size and granularity) on the energy resolution. Our optical MC simulations predict only a marginal deterioration of the spectroscopic performance for pixels of 3×3 mm$^2$.

---

[*] Present address: IMDEA Nanociencia, Campus de Cantoblanco, 28049 Madrid, Spain
[†] Corresponding author.

# Contents

## 1. Introduction and motivation

Advances in both radiation detectors and neutron-beam facilities have led, over the last decades, to the discovery of many new facets of slow (s-) process nucleosynthesis and to a progressive refinement of theoretical models of stellar and galactic chemical evolution [1]. One of the approaches most commonly used for time-of-flight neutron-capture measurements consists on the use of low efficiency radiation detectors in combination with the so-called pulse-height weighting technique (PHWT) [2][5]. For this technique, any kind of radiation detector may be used, in so far as the gamma-ray detection probability remains low enough to avoid registering two- or more gamma-quanta from the same neutron capture cascade. Such approach has the fundamental advantage of enabling a large flexibility for the design of the detection apparatus itself and thus, the possibility to largely optimize the system in order to minimize the impact of neutron-induced gamma-ray backgrounds. In this respect, a large progress has been made from the first Moxon-Rae detectors developed in the sixties **Error! Reference source not found.**, which were soon afterwards replaced by organic $C_6F_6$ liquid scintillation total-energy detectors [2][4]. Presently, the state-of-the-art in the field is represented by carbon-fibre based $C_6D_6$ liquid-scintillation detectors [6], characterized by a very low intrinsic neutron sensitivity.

However, one of the main background sources in neutron time-of-flight experiments arises from neutrons, which are scattered in the sample under study, and subsequently become captured (prompt or after thermalization) in the surrounding materials, such as walls, structural elements, etc. This is nicely illustrated in Fig.6 of [7], which shows that gamma-rays from neutron capture in the walls of the experimental area represent the main background limitation in the relevant energy range for astrophysics, between 1 keV and 100 keV. To a large extent, this has been a constraint in recent (n,γ) experiments, particularly those involving small amounts of radioactive samples [8][9], a situation which has led to a limited astrophysical interpretation of the corresponding branching nuclei [10].

With the aim of exploring new ways of improving this situation, in the framework of the ERC-funded project HYMNS [11], we are investigating the possibility of using low-efficiency radiation detectors with gamma-ray imaging capability, so-called i-TED[12], in combination with the PHWT. Building on our previous successful experience with a mechanically collimated gamma-camera [13], we are now developing an i-TED prototype in order to perform first proof-of-principle measurements at CERN n_TOF [13]. The i-TED detection system is based on the Compton scattering law and thus, high energy- and position-resolution is required in order to reconstruct, on an event-by-event basis, the Compton cone of possible incident radiation directions, as described in [14]. Since the gamma-ray efficiency of such an apparatus is very low ($\varepsilon_\gamma \ll 1$), we intend to develop an i-TED array based on several Compton modules around the target, in order to cover a larger solid angle around the capture sample.

In terms of position and energy response, several recent experimental studies have reported promising results for lanthanum halide crystals coupled to silicon photomultipliers (SiPMs). In particular, Compton devices developed for gamma-ray astronomy such as ASCOT [15], targeting a similar energy range as the one involved in neutron capture reactions (0.5-10 MeV), have found energy resolutions of 5.3% FWHM at 662 keV using a cubic $26\times26\times26$ mm$^3$ LaBr$_3$ crystal coupled to an 8x8 pixel SiPM array [16]. In [17] a resolution of 4% FWHM was found for a $28\times28\times20$ mm$^3$ LaBr$_3$ crystal with $8\times8$ pixel SiPM readout. The later work also reports spatial resolutions of 2.9 mm FWHM in the transversal XY plane and 5.2 mm FWHM for the depth of interaction (DOI) for a 10 mm thick CeBr$_3$ crystal. In the field of nuclear medicine, there are promising results [18] for large monolithic LaBr$_3$ crystals of $50\times50\times30$ mm$^3$ using a segmented photomultiplier-tube (PMT) and analogue readout electronics, for which energy resolutions of 3.8% FWHM and spatial resolutions of 5.5 mm FWHM were found. Smaller crystals of $16\times18\times5$ mm$^3$ coupled to a 4x4 pixel SiPM have led to resolutions of around 6% FWHM [19]. Albeit auspicious, none of these studies cover the i-TED project needs of high-efficiency large monolithic crystals with SiPM readout.

In this article, we explore the spectroscopic performance of large area ($50\times50$ mm$^2$) monolithic LaCl$_3$(Ce) crystals of several thicknesses, from 10 to 30 mm, coupled to pixelated silicon photomultipliers. Such kind of position sensitive detectors (PSDs) will be the main building elements of the afore mentioned i-TED prototype. For our application, on one side, a thin pixel granularity leads to a more accurate sampling of the scintillation light-distribution, which might lead to a more accurate spatial response. The spatial performance of such PSDs is out of the scope of the present article, and it will be reported on a separate work. On the other hand, the higher dead-space related to the high granularity also implies a loss of scintillation photons and thus, a deterioration of the energy response. Therefore, a pixel granularity needs to be found, which allows for the best trade-off between energy- and position-resolution, while keeping under reasonable levels the scalability and complexity of the system in terms of readout channels.

In order to study these aspects, we have carried out a thorough Monte Carlo (MC) study of the PSDs themselves. The latter includes both electromagnetic physics and optical interactions for all the scintillation-photon histories generated at each gamma-ray interaction inside the crystal. This allows us to study, in a realistic fashion, the energy- and position-response of the PSDs. The results obtained from these calculations are experimentally validated by means of a series of benchmark measurements carried out in the laboratory, both with a conventional mono-cathode PMT and with an 8×8 pixel SiPM. After validation of our computation model, we use it to investigate the impact of granularity on the spectroscopic performance of the PSD.

In Section 2 we describe the experimental apparatus and the results obtained for the energy resolution measurements at 662 keV, both for a conventional PMT used as reference and for each crystal coupled to the 8×8 channels SiPM. A general description of the MC calculations and a comparison of the results obtained in the simulation versus the laboratory measurements is given in Section 3. Once the MC model is validated, we use it to study the impact that a different pixel granularity would have on the spectroscopic performance of the detector. A summary and outlook of our results is provided in Section 4.

## 2. Experimental apparatus, methodology and spectroscopic performance

### 2.1. Scintillation crystals and photosensors

We have carried out measurements with three $LaCl_3(Ce)$ crystals, all of them with a square size of 50×50 mm$^2$ and different thicknesses of 10, 20 and 30 mm. Reference measurements were made by coupling each $LaCl_3(Ce)$ crystal to a square-photocathode Hamamatsu R6236 PMT. A homogenous optical contact was achieved by using silicone grease (BC-630) between the optical window of the crystal and the photocathode. This PMT features 8 dynode stages, a photocathode area of 54×54 mm$^2$ and a typical quantum efficiency of 30% at peak wavelength.

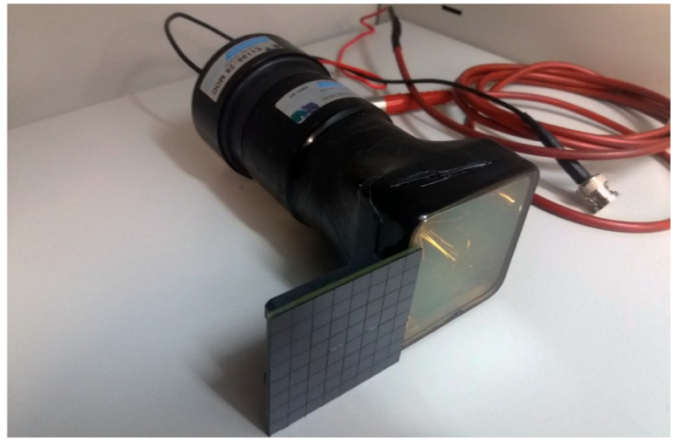

**Fig. 1: Comparative picture of the SiPM array (left) and the monocathode photomultiplier tube (right).**

The SiPM used was a large area array from SensL (ArrayJ-60035-64P-PCB). A comparative picture of the SiPM with respect to the PMT is shown in Fig.1. This SiPM has 8×8 channels distributed over a PCB with a size of 50.4×50.4 mm$^2$, with a pixel pitch of 6.33 mm and a fill-factor of 75%. The quantum efficiency of the silicon sensor depends on the depletion voltage value, ranging from 35% up to 50% at peak wavelength for voltages over the breakdown value between 2.5 V and 6 V, respectively. Optical grease BC-630 was also used for the coupling between the SiPM and the scintillation crystal.

## 2.2. Power and readout electronics

A Tennelec TC954 high-voltage module was used to power the Hamamatsu PMT at its nominal value of -580 V. The current signal from the last PMT dynode was fed into a CANBERRA-2005 preamplifier and shaped by means of a Tennelec TC-244 amplifier. The analogue output from the latter was used to obtain the pulse-height spectra by means of a multi-channel analyzer (Palmtop MCA8k-01). These spectra were exported in ASCII format and afterwards analyzed using the CERN ROOT package [24] to derive the energy resolution (see below).

In order to acquire data with the SiPM, we plugged it into a breakout sum-board (ArrayX-BOB6-64S) provided also by the same manufacturer. Such a PCB is designed to merge all the standard pixel anode signals from the SiPM and thus provides a charge-signal proportional to the total number of scintillation photons detected with the SiPM. The SiPM was biased at 5 V beyond the nominal breakdown voltage of 25 V by using a GRELCO GVD305SF voltage-supply unit. The summed anodes signal was then dc-decoupled by means of a 10 nF capacitor and fed to the Tennelec TC-244 amplifier. The shaped output signal was fed into the MCA for getting the corresponding pulse-height spectrum.

## 2.3. Spectroscopic performance: energy resolution measurement at 662 keV

Each combination crystal-PMT and crystal-SiPM was calibrated in energy by means of dedicated $^{22}$Na, $^{137}$Cs and $^{60}$Co measurements in the energy range from 511 to 1332 keV. Additionally, a background measurement was carried out in order to subtract the ambient and the intrinsic crystal (α-) background contributions from the source spectrum. A picture of the set-up used for calibration and characterization measurements using the SiPM is displayed in Fig.2.

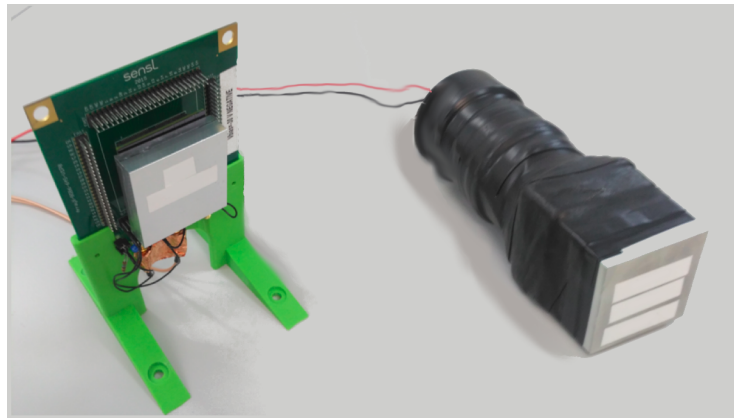

**Fig. 2: Set-up used for the characterization of the LaCl$_3$(Ce) crystal of 10 mm thickness coupled to the SiPM (left) and the LaCl$_3$(Ce) crystal of 30 mm thickness coupled to the PMT (right).**

In order to accurately determine the energy resolution for each crystal-photosensor assembly we have implemented an algorithm, which performs a least-squares minimization between the measured pulse-height spectrum and the one calculated from electromagnetic interactions of the $^{137}$Cs beta-decay using the Geant4 simulation code (see Section 3). The simulated spectrum, initially with an ideally narrow resolution, is convoluted during the minimization process with a Gaussian distribution until it matches nearly pefectly the experimental one (see Fig.3). Hereby, the energy dependency of the Gaussian width is assumed to have a linear dependency with the square-root value of the deposited energy [20]. We have found this approach significantly more reliable and accurate than the more commonly used method [21] of partially fitting a single Gaussian distribution to the full-energy peak in the experimental spectrum. The energy calibrated pulse-height spectra for the three $LaCl_3(Ce)$ crystals are shown in Fig.3 for both PMT and SiPM readout schemes employed. Gamma-ray events leading to full-energy deposition show, in both cases, a very similar signature, which reflects their similar photo-detection performance. The main difference arises in the energy range between the upper Compton edge and the full-energy deposition events, where a higher contribution was found for all measurements made with the SiPM.

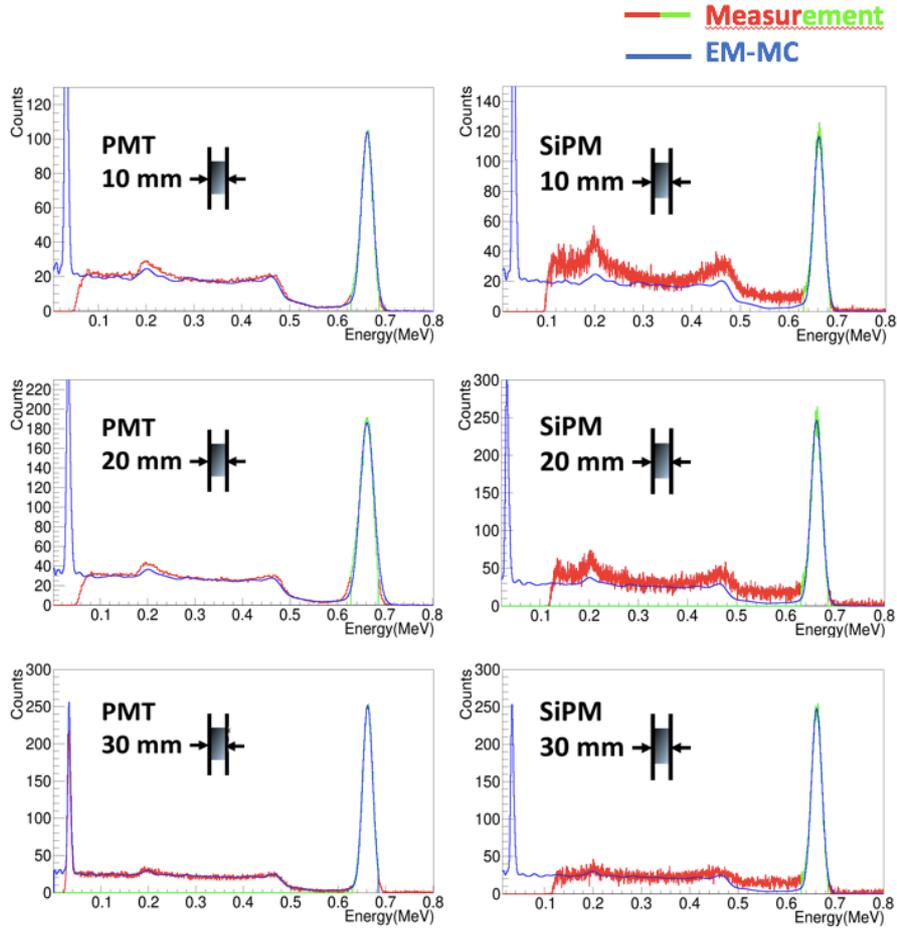

**Fig. 3: Calibrated energy spectra for the $^{137}$Cs source measured with the $LaCl_3$ crystals coupled to PMT (left column) and to the SiPM (right column). From top to bottom the crystal thicknesses are of 10, 20 and 30 mm. The green region of the experimental spectrum shows the energy range chosen for the least-squares minimization to determine the energy resolution.**

Measurements made with the SiPM photosensor yield, on average, a better energy resolution than those carried out with the conventional mono-cathode PMT. Energy resolution FWHM values obtained for the six combinations of crystal-photosensor are shown in Table 1 and plotted in Fig.4.

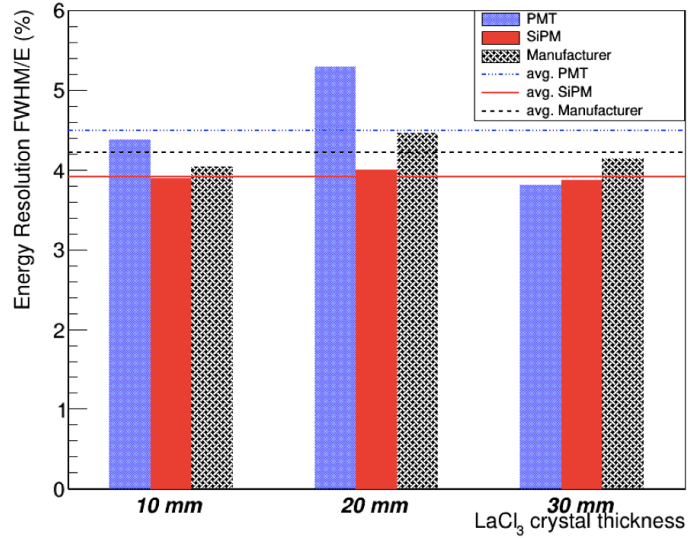

**Fig. 4:** Energy resolution (FWHM) at 662 keV obtained for the three different crystals coupled to the PMT (blue) and to the SiPM (red); solid lines represent their average values. The resolutions given by the manufacturer of the crystals (using a PMT) are also displayed (dashed bars) along with the corresponding average resolution.

| 10 mm | | 20 mm | | 30 mm | |
|---|---|---|---|---|---|
| PMT | SiPM | PMT | SiPM | PMT | SiPM |
| 4.37(2)% | 3.886(5)% | 5.29(2)% | 3.992(4)% | 3.803(5)% | 3.867(4)% |

**Table 1.** Energy Resolutions (FWHM) at 662 keV obtained for the three different crystals coupled to the PMT and to the SiPM

On average, the resolution obtained with SiPM readout is 3.92% FWHM, to be compared with the 4.49% FWHM found for the measurements with the PMT. Thus, maximum variations of about 0.8% and 0.07% are found for individual crystal-photosensor configurations with respect to the average resolution value for PMT and SiPM, respectively (see Fig.4). This better spectroscopic performance found here for SiPM with respect to PMT is at variance with the comparison reported previously, for example in [17], for PMT and SiPM sensors. This result may well be ascribed to the higher quantum photo-detection efficiency and fill factor of the new generation of SiPMs. This aspect will be further discussed in the section below on the basis of detailed MC simulations that include the optical transport and absorption of scintillation photons in the scintillation crystal and in the readout photosensor. Apart from the differences in the average values for the energy resolution, no clear systematic trend has been found regarding the thickness/size aspect-ratio of the crystals.

Two conclusions can be derived from these measurements, which are important for the future development of i-TED. Firstly, $LaCl_3(Ce)$ crystals, which are preferred in our case with respect to LaBr3(Ce) due to the lower neutron capture cross section of Chlorine compared to Bromine[12], can provide an energy resolution similar to that of $LaBr_3$ scintillators[17][18][19]. Secondly, the good performance of such crystals in terms of energy resolution does not fade when replacing a mono-cathode PMT by a pixelated SiPM. On the contrary, it improves in relative terms by about 10%. Given the ongoing progress on Si-photosensor technology, where one can envisage enhancement in both photo-detection efficiencies and fill factors, one may even expect that this trend leads to even better spectroscopic performances in the near future.

## 3. Monte Carlo modelling of the experimental apparatus

### 3.1. Implementation in Geant4 multi-thread

The Geant4 version 10.3 simulation software [22] has been used to model our experimental set-up and, in particular, to develop a toolkit to study the impact of the SiPM pixel size on both the spatial- and energy-response function of our system. The material and geometrical description of our detector includes the $LaCl_3$ scintillation crystal, a 100 μm thin layer of air, a diffusive reflector made from Teflon, the aluminum encapsulation and the optical window as shown in Fig.5.

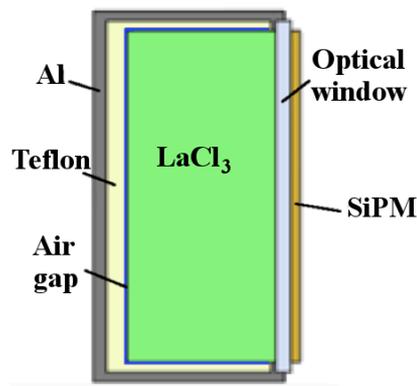

**Fig. 5: Schematic view of the $LaCl_3$-SiPM detector as implemented in the MC simulation (left). Example of one event with all the secondary scintillation photon histories displayed (right).**

The use of the optical capabilities in Geant4 requires, in addition to the optical physics module, the accurate definition of the optical properties for all materials and interfaces involved in the simulation. All these properties have been included in our calculation as a function of the photon momentum, covering the wavelength range from 300 up to 600 nm, with a binning resolution of 3 nm.

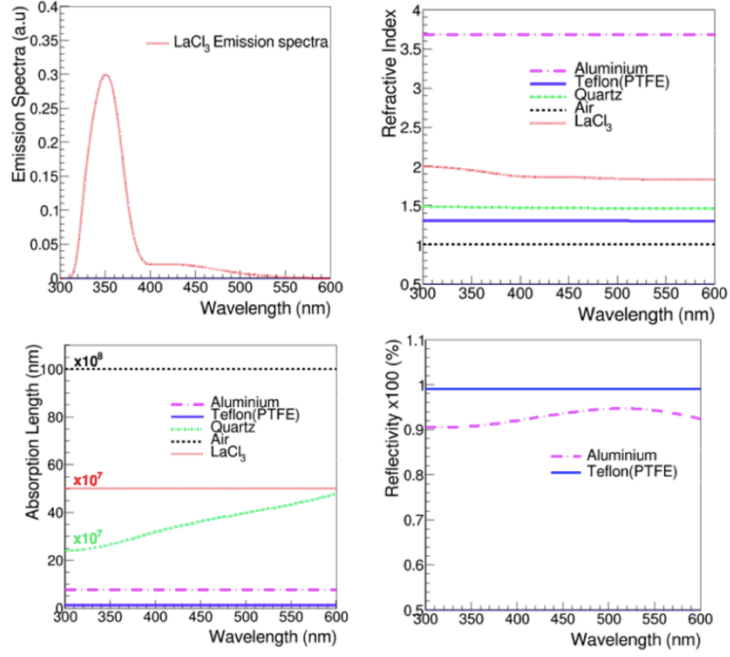

**Fig. 6: Numerical distributions of the main optical properties as implemented in the simulation: scintillation light distribution for LaCl$_3$(Ce) (top-left), refractive index for the materials of our detectors (top-right), absorption lengths (bottom-left) and reflectivity of aluminum and Teflon (bottom-right). See text for details.**

For the sensitive detection volume, a scintillation yield is provided (Fig. 6), which accounts for the number and momenta of photons produced by the ionizing radiation per keV of deposited energy. The scintillation spectrum is included as a function of the photon wavelength ($\lambda$), which was provided by the crystal manufacturer. The definition of the scintillation process involves also the decay time within the fast scintillation component, the yield-ratio or portion of photons emitted via the fast component and a fudge factor called resolution scale, which affects the statistical distribution of generated photons (see Table 2).

| Crystal Thickness (mm) | Scintillation yield (ph/keV) | Decay Time (ns) | Yield Ratio (%) | Resolution Scale |
|---|---|---|---|---|
| 10 | 48 | 28 | 100 | 1.7 |
| 20 | 48 | 28 | 100 | 2.8 |
| 30 | 48 | 28 | 100 | 1 |

**Table 2. Values of the properties used to define the LaCl$_3$ crystal sensitive volume in the simulation. See text for details.**

Materials involved in the optical processes are characterized by their refractive index, absorption length and, for reflecting materials, reflectivity (Fig. 6). The quantum efficiency of the photosensor was also modelled according to the data provided by the corresponding manufacturer. Such distributions are shown in Fig.7 for the PMT and the SiPM.

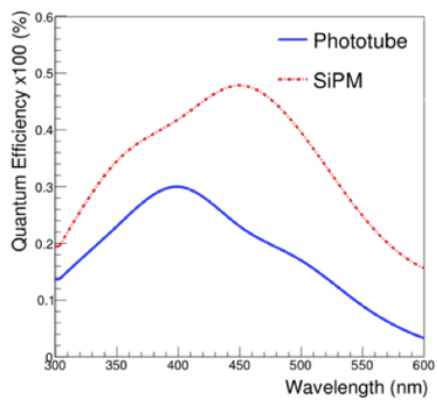

**Fig. 7:** Quantum efficiency for the two photosensors implemented in the code.

| Material | Refractive Index (@ 350nm) | Absorption Length (@ 350nm) | Reflectivity (@ 350nm) |
|---|---|---|---|
| $LaCl_3$ | 1.8243 | 50 cm | - |
| Air | 1.0028 | 1000 cm | - |
| Teflon | 1.3031 | 1 nm | 99% |
| Aluminum | 3.6744 | 7.6 nm | 94% |
| Quartz | 1.4599 | 43.3 cm | - |

**Table 3.** Optical properties defined for the materials present at the simulation. The values shown are referred to the momentum that most of the photons carry (350 nm).

Two different approaches are available in Geant4 to model the reflection and refraction processes of the scintillation quanta. The *glisur* model applies directly the law of Snell for an incoming photon impinging on a surface. In this case, a rough surface is considered to be a collection of *microfacets*, whose normal vectors are the combination of the normal vector for the average surface and a vector obtained with a random point contained in a sphere of certain radius. The latter is given by a free parameter, which is related to the polish-level of the crystal. Alternatively, the *unified* model [23] distributes the micro-facets orientation following a Gaussian distribution and photons will undergo a specular reflection in this surface together with other contributions such as backscattering and Lambertian reflection. The unified model requires a surface characterization and a detailed knowledge of the mentioned contribution probabilities, which were not available for the present work. For this reason, we used the *glisur* model in our simulations. As it is demonstrated below, the glisur model was indeed found sufficiently accurate for a fair reproduction of the measured spectroscopic response. Using this model, one can account for surfaces with ground-finish (which lead to a perfect diffuse Lambertian reflection), for surfaces with polished finish (which yield specular reflection) or for a linear combination of them. The optical window of the $LaCl_3$ scintillation crystals used in this work had a polished surface and a rough finish on the other five surfaces (base and walls). However, the degree of roughness was not available and, for this reason, we adjusted the polish-level within the *glisur* model in order to account for this unspecified property. We found a polish level of 0.7 convenient for a reasonable reproduction of our measured spectra, thus indicating a specular rather than diffusive situation. Other feature that must be established for the surface definition is the kind of transition between materials, which can be defined as a *dielectric-to-dielectric* transition or *dielectric-to-* *metal*, being the latter the one chosen to define any reflecting surface regardless of its electrical properties. In order account for boundary processes the surface finish needs to be properly defined. In the present version of the code, one can choose between *polished, polishedfrontpainted, polishedbackpainted, ground, groundfrontpainted* and *groundbackpainted* surface finish. The surfaces implemented in our calculation along with the values of the parameters defined are listed in Table 4.

| Surface | Surface Type | Surface Finish | Surface Polish Level |
|---|---|---|---|
| *LaCl3 - Air* | *Dielectric - Dielectric* | *Ground* | 0.7 |
| *LaCl$_3$ – Quartz* | *Dielectric - Dielectric* | *Polished* | - |
| *Air – Teflon* | *Dielectric - Metal* | - | - |
| *Air – Aluminum* | *Dielectric - Metal* | - | - |
| *Air – Quartz* | *Dielectric - Dielectric* | *Polished* | - |
| *Teflon – Aluminum* | *Dielectric - Metal* | - | - |
| *Teflon – Quartz* | *Dielectric -Metal* | - | - |
| *Aluminum – Quartz* | *Dielectric - Metal* | - | - |

**Table 4. Surface properties defined for the interfaces between the materials conforming the detector.**

Finally, we have made use of the possibility of using the Geant4 multi-thread mode in combination with a multi-core computer. This allows us to execute in parallel separate Geant4 threads concurrently by separate hardware threads, thus enhancing remarkably the processing capability and keeping the total amount of CPU-time within reasonable limits. The efficiency of the parallelization option is demonstrated in Fig.8, which displays the results from a series of MC computations covering the range from $1\times10^6$ up to $3\times10^8$ scintillation photons. In each MC run, a total number of 1, 2, 4 and 8 threads were used. Also in Fig.8 we show the average CPU time needed for each simulated optical photon as a function of the number of parallel threads. Thus, one can conclude that, increasing the number of threads beyond 8 does not contribute to a significant reduction of CPU time. For this reason, we carried out all the MC calculations presented in this article with 8 threads using an Intel i7 processor.

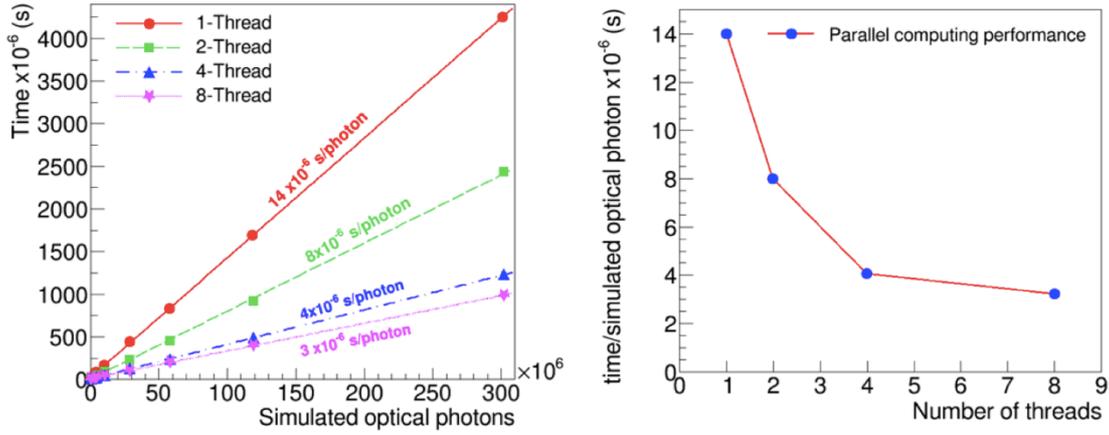

**Fig. 8:** CPU computation time as a function of the number of simulated optical photons (left). Average CPU computation time required for each optical photon, as a function of the number of threads (right).

For the study and analysis of the simulated data the CERN ROOT analysis toolkit [24] was used. Geant4 implements an analysis manager (G4RootAnalysisManager) that allows us to store easily all the desired information on a ROOT formatted file.

### 3.2. Energy resolution: MC results and comparison versus measurements

In order to have a direct comparison between the number of photons detected in the photosensor -as predicted by the simulation- and the measured spectra, the simulation has been scaled on the horizontal axis to translate the number of detected photons into calibrated energy-units. As demonstrated in Fig.9, the optical simulation of the response function yields a fairly good agreement with the measured spectra, both for PMT and SiPM. The response function simulated in all three cases for the PMT-readout shows a nearly perfect agreement with the measured spectra and, a significant improvement with respect to the EM-MC simulation (see Fig.3). Regarding the SiPM results, the agreement with the experimental response function is worse than for the PMT. This fact was expected as the aforementioned fudge factor (see Table 2) was defined for the better reproduction of the PMT measurements. However, SiPM results using the Optical MC are substantially better than the results obtained from the EM-MC calculation shown before in Fig.3. This result reflects the impact of the SiPM features (dead-zones) on the photon-counting process, and the need of simulating the optical part for a reliable description of the response function when using a pixelated SiPM. In particular, the balance between full-energy and Compton events is better reproduced by the new simulations, being the main discrepancy the aforementioned continuum between the upper Compton border and the full energy peak.

For each crystal-photosensor assembly, the ideal electromagnetic MC response is convoluted with a Gaussian distribution until it fits the new optical MC response. In this way, the width (FWHM) of the full-energy peak in the spectrum obtained from the Optical MC (photon-units) can be directly used to obtain the energy resolution. The results displayed in Fig.10 and reported in Table 5 indicate a better agreement between simulation and measurements for the PMT with respect to the SiPM. In the former case the largest deviation (2% in relative terms) is found for the 30 mm thick crystal. Regarding the SiPM, differences between the simulation of optical-photons and the spectroscopic measurements are larger for the 20 mm and 30 mm thick crystals, with discrepancies of +20% and -13%, respectively. The relatively large deviations found for the SiPM can be ascribed to the approach followed in Sec.3, where the unknown fudge-factors were adjusted within the MC simulations to reproduce the (simpler) set-up of each crystal coupled to the monocathode PMT. Therefore, it is reasonable that the results derived for the optical SiPM simulations are appreciably biased by this (PMT-based) fudge-factor.

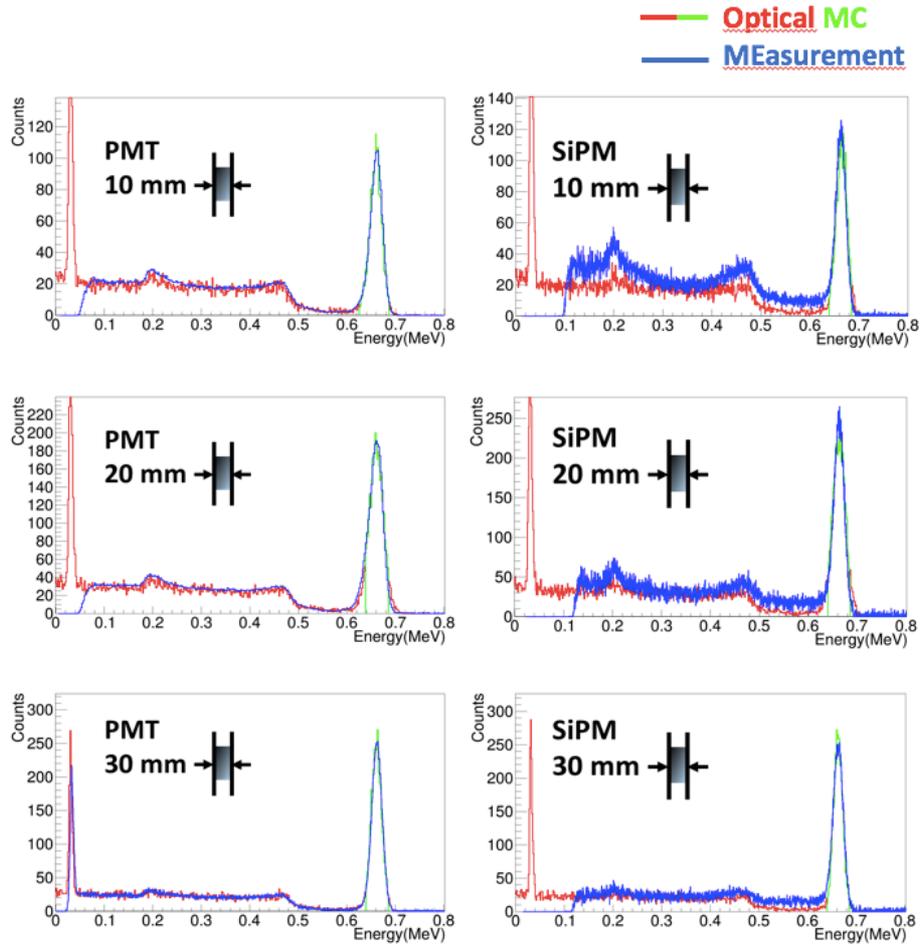

**Fig. 9: Optical MC simulation of the response function (red-green spectra) for the PMT readout (left column) and SiPM (right column). From top-to-bottom the panels show the results for detectors with crystal thickness of 10, 20 and 30mm.**

On average, the optical MC simulation for the energy resolution of detectors with SiPM readout (4.0% FWHM) is in fairly good agreement with the average of the measured values (3.9% FWHM). A similar agreement is found for the comparison between the average of the optically simulated resolution for the PMT (4.49% FWHM) and the measurements (4.50% FWHM).

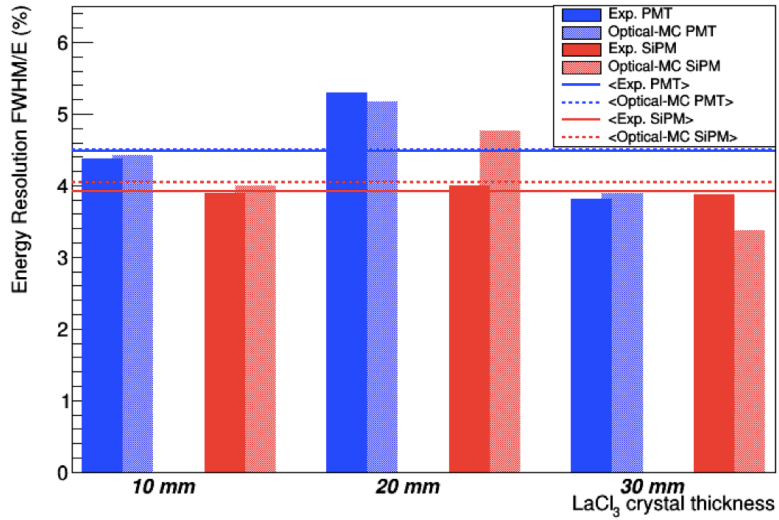

**Fig. 10:** Comparison of experimental energy resolution (bold) versus optical MC simulated values (light).

|  | 10 mm | | 20 mm | | 30 mm | |
|---|---|---|---|---|---|---|
|  | PMT | SiPM | PMT | SiPM | PMT | SiPM |
| FWHM/E | 4.428(12)% | 4.003(8)% | 4.77(2)% | 5.181(8)% | 3.89(7)% | 3.369(5)% |
| Δ(Exp./MC) | 1.4% | 3.0% | -2.2% | 19.4% | 2.3% | -12.9% |

**Table 5.** Energy resolution (FWHM) at 662 keV determined by means of Optical MC simulations for the three different crystals coupled to the PMT and to the 6×6 mm$^2$ pixelated SiPM. In the last row, the difference in relative terms between the estimated resolution and the measured one is listed.

As shown in Table 5 and looking now at the maximal individual differences, our MC model allows us to estimate within ±19% (±3%) in relative terms the energy resolution for large monolithic LaCl$_3$ crystals optically coupled to pixelated SiPM (PMT).

Encouraged by this result, we use our code in order to infer the expected performance of a SiPM with a thinner pixel-size of 3×3 mm$^2$. For this calculation, we use realistic technical values from commercially available SiPM, in particular those from the sensL ArrayJ-30035-64P-PCB. The idea behind this simulation resides on the fact that, although less scintillation photons will be registered due to increasing dead-areas, an enhancement in position reconstruction might be achieved on XY and/or DOI, due to the thinner sampling resolution. As mentioned before, the spatial response of these PSDs will be the focus of a forthcoming article. For determining the energy resolution, we have followed the same methodology previously described. The ideal response function simulated by means of the EM-MC calculation was convoluted with a Gaussian distribution with a square root dependency on the energy, and the response function from the optical MC simulation was converted to energy units by means of a linear relationship. Thus, the accurate value for the energy resolution of the optical simulation is found for the best agreement between both simulations, as shown in Fig.11.

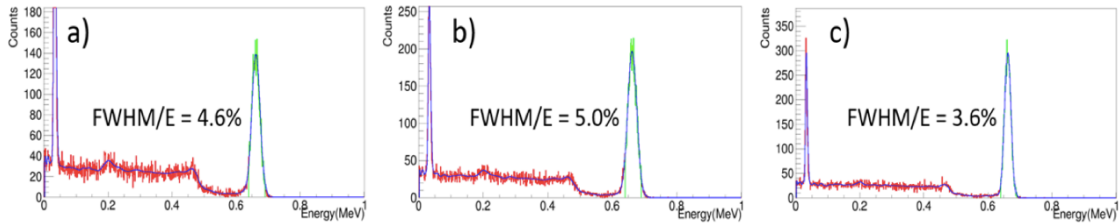

**Fig. 11:** Optical MC simulation (red-green spectrum) of the response function for the three crystals with thicknesses of 10 (a), 20 (b) and 30 mm (c), each readout with a 16×16 channel SiPM with a pixel size of 3×3 mm$^2$. The blue spectra represent the electromagnetic MC simulations. The green region of the Optical MC spectra represents the energy interval used for the least squares minimization. See text for details.

|  | 10 mm | 20 mm | 30 mm |
|---|---|---|---|
| FWHM/E | 4.60(2)% | 4.96(1)% | 3.595(5)% |
| Δ(3mm/6mm) | 14.9% | 4.9% | 6.9% |

**Table 6.** Energy resolution (FWHM) at 662 keV estimated by means of Optical MC simulations for the three LaCl$_3$ crystals coupled to a 3×3 mm$^2$ pixelated SiPM. The bottom row shows the difference in relative terms between the estimated resolution for the 3×3 mm$^2$ SiPM and the 6×6 mm$^2$.

The results reported in Table 6 for the 3×3 mm$^2$ pixel-size SiPMs are, at first sight, somewhat surprising, as they indicate that the energy resolution is only marginally affected by the thinner pixel-size, despite of the significantly larger dead-space, when compared to the 6×6 mm$^2$ pixel (see Table 6). This effect can be understood by the fact that the loss in resolution is proportional to the square-root value of the ratios between the dead-zones for the low- and high-granularity SiPMs, which is much smaller than the simple ratio of the dead-zones for each SiPM.

In summary, there are essentially two aspects to consider regarding the usefulness of the 3×3 mm$^2$ pixel size for our application in i-TED [11]. Firstly, the higher cost and complexity in readout- and processing electronics, which for a certain detection area need to be scaled by a factor of four in order to use 3×3 mm$^2$ pixels, when compared to the instrumentation required for the 6×6 mm$^2$ pixels. Secondly, the performance of the thinner pixelation in terms of spatial sensitivity, both along the XY-plane and the DOI. In order to evaluate the impact of the pixelation granularity on the performance of the spatial reconstruction we have carried out dedicated measurements and explored different position reconstruction algorithms, which will be reported in a separate paper.

## 4. Summary and outlook

We have developed large monolithic position-sensitive lanthanum halide detectors and accurately characterized their spectroscopic performance. With a total sensitive surface of 50×50 mm$^2$, they are of the largest reported in the literature using SiPM readout. These detectors are primarily intended for the deployment of a total-energy detector with imaging capability for radiative neutron capture experiments at TOF facilities. Using the latest generation of high-quantum efficiency and high fill-factor SiPMs, we were able to obtain a rather good energy resolution, with an average value of 3.92% FWHM for crystal thicknesses of 10, 20 and 30 mm. We have also quantitatively explored the energy response of our apparatus, finding similar capabilities for both 3×3 mm$^2$ and 6×6 mm$^2$ pixel sizes. The developed MC code allows us to reproduce fairly well the spectroscopic and spatial behaviour of our detectors. According to our calculations a thinner pixel size of 3×3 mm$^2$ is not expected to significantly decrease the energy resolution, thus representing only an advantage in case that the spatial performance of the thinner pixellation turns out to be substantially superior to that of the 6×6mm$^2$ pixels. These aspects are presently under study in our group and will be reported in a future publication.

## Acknowledgments

This project has received funding from the European Research Council (ERC) under the European Union's Horizon 2020 research and innovation programme (grant agreement No. 681740).
We acknowledge support from the Spanish project FPA2014-52823-C2-1-P. We acknowledge support from IFIC's Electronics Workshop, in particular the help of M. Lopez Redondo, as well as the help from M. Monserrate Sabroso at the Mechanics Workshop of IFIC. We acknowledge A. González and Liczandro Hernández, from I3M at UPV for helpful discussions.